\documentclass[twocolumn,aps,amsmath,amssymb,floatfix]{revtex4}
\usepackage{graphicx}
\usepackage{dcolumn}
\usepackage{bm}
\usepackage{setspace}

\begin{document}


\title{Accuracy of direct gradient sensing by single cells}

\author{Robert G. Endres$^{1,2,3}$ and Ned S. Wingreen$^{1}$}
\affiliation{
$^{1}$Department of Molecular Biology, Princeton University, Princeton, NJ 08544-1014, USA\\
$^{2}$Division of Molecular Biosciences, Imperial College London, London SW7 2AZ, United Kingdom\\
$^{3}$Centre for Integrated Systems Biology at Imperial College, Imperial College London, 
London SW7 2AZ, United Kingdom\\
}


\begin{abstract} 
Many types of cells are able to accurately sense shallow gradients of
chemicals across their diameters, allowing the cells to move towards or away
from chemical sources. This chemotactic ability relies on the remarkable capacity
of cells to infer gradients from  particles randomly
arriving at cell-surface receptors by diffusion. Whereas the physical limits
of concentration sensing by cells have been explored, there is no theory for
the physical limits of gradient sensing. Here, we derive such a theory, using
as models a perfectly absorbing sphere and a perfectly monitoring sphere,
which, respectively, infer gradients from the absorbed surface particle
density or the positions of freely diffusing particles inside a spherical
volume. We find that the perfectly absorbing sphere is superior to the
perfectly monitoring sphere, both for concentration and gradient sensing,
since previously observed particles are never remeasured. The superiority of
the absorbing sphere helps explain the presence at the surfaces of cells of signal degrading enzymes,
such as PDE for cAMP in {\it Dictyostelium discoideum} (Dicty) 
and BAR1 for mating factor $\alpha$ in {\it Saccharomyces cerevisiae} (budding yeast).
Quantitatively, our theory compares favorably to recent measurements of Dicty moving up a cAMP gradient, 
suggesting these cells operate near the physical limits of gradient detection.
\end{abstract}


\maketitle

Cells are able to sense gradients of chemical concentration 
with extremely high sensitivity. This is done either directly, by measuring spatial gradients
across the cell diameter, or indirectly, by temporally sensing gradients while 
moving. In temporal sensing, a cell modifies its swimming behavior according to whether 
a chemical concentration is rising or falling in time \cite{berg99}. 
This mode of sensing is typical of small, fast moving bacteria such as 
{\it Escherichia coli}, which can respond to changes in concentration as low as 
3.2 nM of the attractant aspartate \cite{manson03}. In contrast, direct spatial sensing is 
prevalent among larger, single-celled eukaryotic organisms such as the slime mold {\it Dictyostelium discoideum} 
(Dicty) and the yeast {\it Saccharomyces cerevisiae} \cite{arkowitz99,manahan04}.
Dicty cells are able to sense a concentration difference of only 1-5\% across the cell \cite{mato75},
corresponding to a difference in receptor occupancy between front and back of only 
5 receptors \cite{haastert07}. Spatial sensing is also performed with high accuracy by cells of the immune system 
including neutrophils and lymphocytes \cite{zigmond77}, as well as by growing synaptic cells and tumor cells. While there
has been great progress in understanding the limits of concentration sensing and signaling in bacteria such as 
{\it E. coli} \cite{berg77,bray98,bialek05,mello05,keymer06,endres06}, very little is know about the 
theoretical limits of direct gradient sensing by eukaryotic cells.

In a recent set of experiments, van Haastert and Postma \cite{haastert07} measured
the Chemotactic Index of Dicty cells in a cAMP gradient (Fig. 1, symbols). 
Chemotactic Index is defined as the distance moved in the direction of the cAMP gradient 
divided by the total distance moved. To obtain the data in Fig. 1, van Haastert and Postma used a pipette containing 
different concentrations of cAMP. Diffusion of cAMP out of the pipette established
a steady-state cAMP gradient, with magnitude a function of distance from the pipette. Chemotaxis was
observed for cells as far as 700 $\mu$m from a pipette filled with $10^{-4}$ M cAMP, 
corresponding to a mean concentration of 7 nM and a gradient of only 0.01 nM/$\mu$m. This remarkable chemotactic ability 
raises the question -- how closely does gradient sensing by Dicty cells compare with the fundamental limits 
on gradient sensing set by diffusion?  
 
Here we derive the fundamental limits of gradient sensing using two models for cells: a perfectly absorbing 
sphere and a perfectly monitoring sphere. 
Within the theory, gradients are estimated by comparing the discrete positions of particles, either absorbed 
on the surface of the sphere or measured inside the spherical volume, 
with the expected continuous distribution originating from a particular gradient. 
We find that a perfectly absorbing sphere 
is superior to a perfectly monitoring sphere for sensing both concentrations and gradients (Table I), since previously 
observed particles are never remeasured. Quantitatively, 
our theory (Fig. 1, solid curves) compares favorably with recent measurements of Dicty cells migrating to a cAMP-filled
pipette \cite{haastert07}, suggesting that chemotactic ability of Dicty approaches the fundamental limits
set by diffusion.


\section{Limits of Concentration Sensing}

In this section, we consider the limits of concentration sensing set by particle diffusion.
Consider as a measurement device a spherical cell of radius $a$ that can measure
the local concentration of a certain dissolved chemical. Such an
idealized device may make measurements following two different strategies: (1) The device can 
either act as a {\it perfectly absorbing sphere} and record the number of absorbed particles on its 
surface or (2) act as a {\it perfectly monitoring sphere} and count the number of particles inside its volume. 
In either case, from the number of particles, an estimate of the chemical concentration can be obtained. However, 
these estimates have an intrinsic uncertainty due to the randomness of particle diffusion.\\

\noindent{\bf Perfectly Absorbing Sphere.}
For the perfectly absorbing sphere the uncertainty in measuring a background chemical concentration $c$ is straightforward
to derive. At steady state, the average particle current impinging on the sphere is $J=4\pi Dac$, where $D$ is the chemical
diffusion constant. The average number of particles absorbed in time $T$ is $N=4\pi DacT$. Since the
particles are independent, $N$ is Poisson distributed, {\it i.e.} $\langle(\delta N)^2\rangle=\langle N\rangle$.
Therefore, the perfectly absorbing 
sphere has a concentration-measurement uncertainty of
\begin{equation}
\frac{\langle(\delta c)^2\rangle}{c^2}=\frac{\langle(\delta N)^2\rangle}{\langle N\rangle^2}
=\frac{1}{4\pi DacT}\label{eq:res_one_c}.
\end{equation}\\

\noindent{\bf Perfectly Monitoring Sphere.} 
The perfectly monitoring sphere was introduced by Berg and Purcell \cite{berg77}
as a parameter-free model for a cell that "perfectly" binds and releases all ligands that contact its surface.
To quantify the time a diffusing particle spends in the cell's vicinity and
is therefore capable of being measured, Berg and Purcell treated the cell as a permeable sphere that
infers the particle concentration by counting the number
of particles $N$ inside its volume, and improves accuracy by averaging over 
several statistically independent measurements.
A simple estimate for the resulting uncertainty in concentration can be obtained as follows: the
number $N$ is Poisson distributed and the cell counts appoximately $N=a^3c$ particles
in its volume at any time. During a time $T$, the cell can make 
$N_\text{meas}=T/(a^2/D)$ independent measurements, where $a^2/D$ is the typical turnover 
time for particles inside the sphere, leading to
\begin{equation}
\frac{\langle(\delta c)^2\rangle}{c^2}=\frac{\langle(\delta N)^2\rangle}{N^2}=
\frac{1}{N_\text{meas}N}\approx\frac{1}{DacT}\label{eq:res_approx}.
\end{equation}

Berg and Purcell \cite{berg77} derived the exact concentration-measurement uncertainty for a 
perfectly monitoring sphere (``perfect instrument'') from the time correlations of
particles inside the sphere, and obtained
\begin{equation}
\frac{\langle(\delta c)^2\rangle}{c^2}=\frac{3}{5\pi DacT}\label{eq:res_two_c},
\end{equation}
which is identical to the estimate in Eq. \ref{eq:res_approx} up to a numerical prefactor.

However, notice that the concentration-measurement uncertainty of the perfectly absorbing sphere is
actually {\it smaller} than that of a perfectly monitoring sphere of the same size, 
because the perfectly absorbing sphere removes particles from the environment, 
and hence does not measure the same particle more than once.


\begin{figure}             
\includegraphics[width=9cm,angle=0]{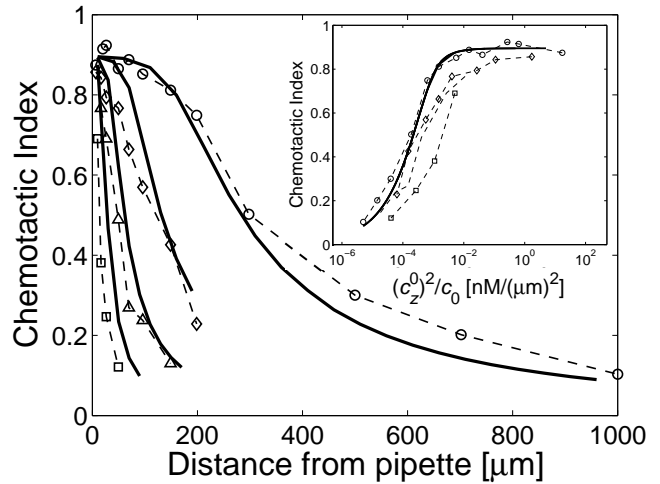}
\caption{Comparison between the Chemotactic Index determined by experiment (symbols and dashed lines) and our theory (solid curves). 
Chemotactic Index is defined as the distance moved by a cell in the direction of a gradient 
divided by the total distance moved.
Experimental data 
was obtained by van Haastert and Postma \cite{haastert07} from {\it Dictyostelium discoideum} 
cells migrating toward pipettes containing four different 
cAMP concentrations, 0.1 $\mu$M (squares), 1 $\mu$M (triangles), 10 $\mu$M (diamonds), 100 $\mu$M (circles). 
The theoretical curves were obtained for a perfectly absorbing sphere using a single fitting parameter 
$Da^3T=1.2\cdot 10^5\ \mu {\rm m}^5$,  
corresponding to, {\it e.g.}, a cAMP diffusion constant of $D=300\  \mu {\rm m}^2/{\rm s}$, a cell radius of 
$a=5\  \mu {\rm m}$, and an averaging time $T=3.2\ {\rm s}$, using the gradient profiles from Ref. \cite{haastert07} and
the Chemotactic Index from Eq. \ref{eq:CI}. 
Experimentally, the Chemotactic Index only reaches approximately 0.9 at zero distance, 
so we rescale our theory curves by 0.9.
Inset: Chemotactic Index as a function of $(c_z^0)^2/c_0$ in units of $n\text{M}/(\mu\text{m})^2$, where $c_z^0$ and $c_0$ are the
gradient and concentration, respectively.}
\label{fig:fig1}
\end{figure}

\begin{figure}             
\includegraphics[width=8.0cm,angle=0]{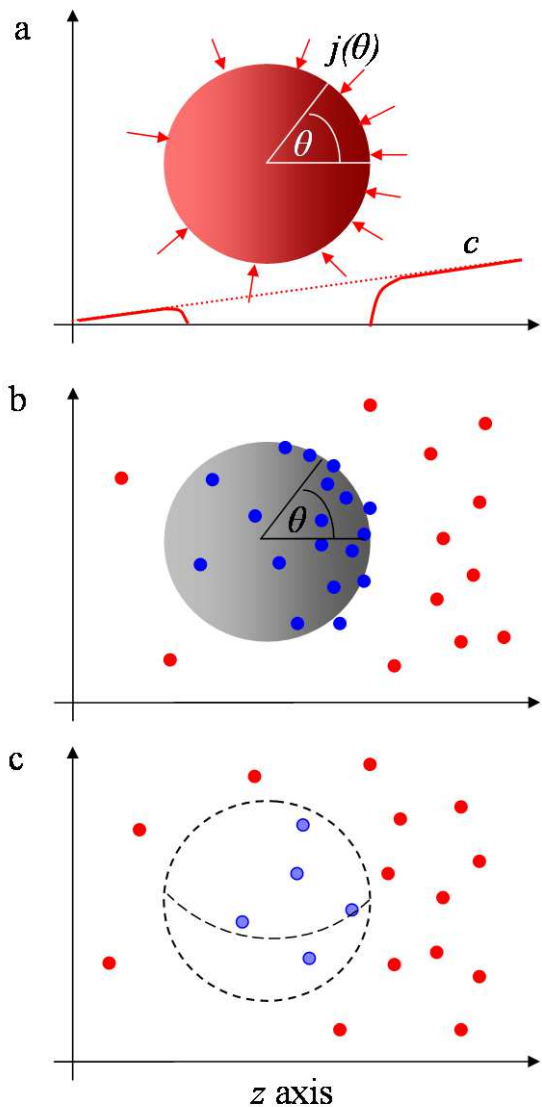}
\caption{Idealized models for gradient sensing by a cell. The gradient points 
along the $z$-axis, which is shown horizontally. (a) Continuum model for a perfectly 
absorbing sphere. The mean particle current density $j(\theta)$ impinging on the sphere
has axial symmetry; $\theta$ measures the angle with respect to the $z$-axis.
At steady-state, the particle concentration $c$ is zero immediately outside
the perfectly absorbing sphere, as shown schematically by the red curve superposed on
the dotted background gradient. (b) Discrete particle model for the perfectly 
absorbing sphere. From the number and positions of particles absorbed during time $T$,
the background particle concentration and gradient can be estimated. (c) Perfectly 
monitoring sphere. Particles diffuse in and out of the sphere without resistance. 
By monitoring, for a time $T$, the number and positions of particles inside the sphere, 
the background concentration and gradient can be estimated.}
\label{fig:fig2}
\end{figure}

\section{Limits of Gradient Sensing}

Now consider the {\it perfectly absorbing sphere} and the {\it perfectly monitoring sphere} 
as devices for measuring the local gradient of a certain dissolved chemical. 
In both cases, measurements of discrete particles can be compared with the expected continuous distribution of particles 
originating from a particular gradient (Fig. 2a), and hence, the gradient can be estimated.
Here, we present in brief a theoretical derivation 
of the intrinsic uncertainty of gradient sensing (for details see supporting information). We find that 
the intrinsic uncertainty is independent of the actual gradient present, and is always much smaller 
(by a factor of 7/60 $\simeq$ 12\%) for the perfectly absorbing sphere.\\

\noindent{\bf Perfectly Absorbing Sphere.}
The average particle current density $\vec{j}=-D\vec{\nabla}c$ impinging on the surface of a
perfectly absorbing sphere of radius $a$ at steady state follows from the 
diffusion equation, $\nabla^2c=0$, and is given in polar coordinates by
\begin{equation}
 j(\theta,\phi)=\frac{Dc_0}{a}+3D\vec c_r\cdot \vec e(\theta,\phi)\label{eq:Jr},
\end{equation}  
where $c_0$ is a constant background concentration, $\vec c_r$ is the background
gradient, and $\vec e(\theta,\phi)=(\cos\phi \sin\theta,\sin\phi \sin\theta,\cos\theta)$ 
is a unit vector normal to the surface of the sphere (see Fig. 2a).
To best estimate the chemical gradient from an observed discrete density of particles absorbed 
at the surface of the sphere during time $T$ (Fig. 2b), requires fitting the observed density 
$\sigma_T^\text{obs}=\sum_{i=1}^{N}\delta(\vec r-\vec r_i)$, where
$N$ is the total number of absorbed particles, to the expected density $j(\theta,\phi)\,T$ 
from Eq. \ref{eq:Jr}.
Since the estimates of the components of the gradient in the $x, y$, and $z$ directions are independent, 
without loss of generality, we consider only the gradient estimate in the $z$-direction, 
{\it i.e.} $c_z=\partial c/\partial z$, and later generalize to an arbitrary gradient. 
From the best fit, the estimate for the gradient in the
$z$-direction after absorption of particles for a time $T$ is given by
\begin{equation}
c_{z}=\frac{\int  \sigma_T^\text{obs}\cos\theta\,dA}{4\pi Da^2T}
=\frac{\sum_{i=1}^{N}\cos\theta_i}{4\pi Da^2T}\label{eq:gradz},
\end{equation}
where $\theta_i$ is the polar angle of the $i$th absorbed particle.
We are  interested in the uncertainty (accuracy) of the gradient measurement, which is given by the variance
\begin{eqnarray}
&&\langle(\delta c_z)^2\rangle=\langle c_z^2\rangle-\langle c_z\rangle^2\label{eq:dcz_with_2}\\
&=&\frac{\langle\sum_{i=1}^{N}\cos^2\theta_i\rangle+
\langle\sum_{i=1}^{N}\sum_{i\neq j}^{N}\cos\theta_i\cos\theta_j\rangle}{(4\pi Da^2T)^2}\nonumber\\
&&\qquad-\frac{\langle\sum_{i=1}^{N}\cos\theta_i\rangle^2}{(4\pi Da^2T)^2}\nonumber\\
&=&\frac{\langle\sum_{i=1}^{N}\cos^2\theta_i\rangle}{(4\pi Da^2T)^2}
=\frac{\langle N\rangle\langle\cos^2\theta\rangle}{(4\pi Da^2T)^2}=\frac{c_0}{12\pi Da^3T}\nonumber.
\end{eqnarray}
The derivation of Eq. \ref{eq:dcz_with_2} made use of the independence of the particles to factorize the expectation value
as $\langle\sum_{i=1}^{N}\sum_{i\neq j}^{N}\cos\theta_i\cos\theta_j\rangle
=\langle N(N-1)\rangle\langle\cos\theta\rangle^2=\langle N\rangle^2\langle\cos\theta\rangle^2$, 
since the number of absorbed particles $N$ is Poisson distributed. 
We also used $\langle N\rangle=4\pi Dac_0T$, as well as
$\langle\cos^2\theta\rangle=1/3$. (The relation $\langle\cos^2\theta_i\rangle = \langle\cos^2\theta\rangle$ 
for absorbed particles holds even in the presence of a true gradient in the $z$ direction 
since the gradient-weighted contribution $~\langle\cos^3\theta\rangle$ is zero.)

In three dimensions, the total uncertainty of the gradient, normalized by $c_0/a$, 
is given by
\begin{equation}
\frac{\langle(\delta c_{\vec r})^2\rangle}{(c_0/a)^2}=
\frac{3\langle(\delta c_{z})^2\rangle}
{(c_0/a)^2}=\frac{1}{4\pi Dac_0 T}\label{eq:res_one},
\end{equation}
with the factor of 3 arising because each component of the gradient contributes independently to the total uncertainty.
This result for the uncertainty in gradient sensing is independent of the magnitude of the actual gradient present,
including the case when no actual gradient is present. Curiously, the result is numerically
identical to the concentration-measurement uncertainty 
(Eq. \ref{eq:res_one_c}).\\

\noindent{\bf Perfectly Monitoring Sphere.}\\
Here, we extend Berg and Purcell's analysis of the perfectly monitoring sphere (``perfect instrument'') 
to include gradient sensing. Specifically, we assume that the monitoring sphere 
measures not only the number but also the positions of all particles in its volume (Fig. 2c).
The best estimate of the gradient is obtained by fitting a concentration 
gradient with a $c=c_0+\vec c_r\cdot\vec r$ 
to the observed time-averaged number density $\frac{1}{T}\int\!dt\rho_\text{obs}(t)=
\frac{1}{T}\int\!dt\sum_{i=1}^{N}\delta(\vec r-\vec r_i(t))$, 
obtained by measuring the exact positions of all the particles inside the volume of the sphere
for a time $T$.
As above we focus on one component of the gradient, namely the gradient in the $z$-direction $c_z$,
and obtain as a best estimate
\begin{equation}
c_{z}=\frac{\frac{1}{T}\int\!dt\int\!dV z\,\rho_\text{obs}(t)}{\int\!dV z^2},
\end{equation}
where the integral $\int\!\!dV$ is over the volume of the sphere.
We are interested in the variance of this estimated gradient
\begin{eqnarray}
 \langle(\delta c_z)^2\rangle&=&\langle c_z^2\rangle-\langle c_z\rangle^2\label{eq:czT2}\\
 &=&\left(\frac{15}{4\pi a^5}\right)^2\left[\langle m_{z,T}^2\rangle
 -\langle m_{z,T}\rangle^2\right]\nonumber,
\end{eqnarray}
where we have used $\int\!\!dV\!\!z^2=4\pi a^5/15$ and have defined
\begin{equation}
m_{z,T}=\frac{1}{T}\int\!dt\int dV z \rho_\text{obs}(t), 
\end{equation}
namely $m_{z,T}$ is the time-averaged total $z$-coordinate of particles 
inside the sphere. The expectation values in Eq. \ref{eq:czT2}
are therefore given by
\begin{eqnarray}
 \langle m_{z,T}^2\rangle&=&\frac{1}{T^2}\left\langle\left(\int\!dt\int\!dV z \rho_\text{obs}(t)\right)^2\right\rangle\label{eq:mzmz}\\
 &=&\frac{1}{T^2}\int_0^T\!dt\int_0^T\!dt'\langle m_z(t)m_z(t')\rangle,\nonumber\\
 \langle m_{z,T}\rangle&=&\frac{1}{T}\left\langle\left(\int\!dt\int\!dV z \rho_\text{obs}(t)\right)\right\rangle\nonumber\\
 &=&\frac{1}{T}\int_0^T\!dt\langle m_z(t)\rangle,\nonumber
\end{eqnarray}
where the quantity $m_z(t)$ is the total of the $z$-coordinates of all the particles inside the sphere at time $t$.
To calculate $m_z(t)$, we consider the sphere embedded inside a much larger volume containing a total of $M$ particles. Then,
$m_z(t)=\sum_{i=1}^{M}z_i(t)$, where $z_i$ is the z-coordinate of particle $i$ if this particle is inside the 
sphere and is zero otherwise. On average there will be $\langle N\rangle=\frac{4}{3}\pi a^3 c_0$ particles inside the sphere
at any time $t$.
The auto-correlation function $\langle m_z(t)m_z(t')\rangle$ of particles inside the sphere at 
time $t$ and time $t'$ can consequently be calculated as
\begin{eqnarray}
 \langle m_z(t)m_z(t')\rangle&=&\langle\sum_{i=1}^M\sum_{j=1}^Mz_i(t)z_j(t')\rangle\label{eq:mzmz2}\\
 &=&\frac{4}{3}\pi a^3c_0\,u(t-t')+\langle m_z(t)\rangle^2\nonumber,
\end{eqnarray}
where we have defined $u(t-t')=\langle z(t)z(t')\rangle$ for a single particle. Substituting Eqs. \ref{eq:mzmz}
and  \ref{eq:mzmz2} into Eq. \ref{eq:czT2} results in
\begin{equation}
\langle(\delta c_z)^2\rangle=\frac{75c_0}{4\pi a^7T^2}\int_0^T\!dt\int_0^T\!dt' u_0(t-t').\label{eq:dcz2}
\end{equation}

By defining a correlation time $\tau_z=(1/a^2)\int_0^\infty\!d\tau u(\tau)$ for the coordinate $z(t)$, the double time integral 
in Eq. \ref{eq:dcz2} can be simplified, provided the time $T$ is much larger than $\tau_z$. Using 
time-reversal symmetry $u(\tau)=u(-\tau)$ for equilibrium diffusion (assuming small gradients), the variance 
simplifies to
\begin{equation}
\langle(\delta c_z)^2\rangle=\frac{75 c_0\tau_z}{2 \pi a^5 T}\label{eq:czT3}.
\end{equation}

The remaining task is to calculate $\tau_z$, the probability that a particle with coordinate $z$ inside the sphere at time $t=0$
is still (or again) inside the sphere at a later time $\tau$. 
We first consider the case in which the background chemical concentration is uniform, and later consider the
presence of an actual gradient.
Based on the solution of the diffusion equation in three dimensions, if a unit amount of chemical is released at 
point $\vec r\,'$, the concentration at point $\vec r$ at a later time $\tau$ is given by
\begin{equation}
f(|\vec r-\vec r\,'|,\tau)=\frac{\exp\left(-\frac{|\vec r-\vec r\,'|^2}{4D\tau}\right)}{(4\pi D\tau)^\frac{3}{2}}.
\end{equation}
Using the result for the time integral from Ref. \cite{berg77},
\begin{equation}
\int_0^\infty\!d\tau\,f(|\vec r-\vec r\,'|,\tau)=\frac{1}{4\pi D|\vec r-\vec r\,'|},
\end{equation}
the correlation time $\tau_z$ can be expressed as a volume integral over the sphere (the initial coordinate
$\vec r\,'$ is uniform in the sphere because we assume a uniform background chemical concentration)
\begin{eqnarray}
\tau_z&=&\frac{1}{\frac{4}{3}\pi a^5}\int_0^\infty\!d\tau\int\!dV\,z\int\!dV'\,z\,'f(|\vec r-\vec r\,'|,\tau)\label{eq:tauz}\\
&=&\frac{3}{16\pi^2 Da^5}\int\!dV\,z\int\!dV'\frac{z\,'}{|\vec r-\vec r\,'|}\nonumber\\
&=&\frac{3}{16\pi^2 Da^5}\int\!dV\,z\,\psi(r,\theta)\nonumber,
\end{eqnarray}
where $r=|\vec r|$. The function
\begin{equation}
\psi(r,\theta)=\int\!dV'\frac{z\,'}{|\vec r-\vec r\,'|}=\int\!dV'\frac{r\,'\cos\theta'}{|\vec r-\vec r\,'|}
\end{equation} 
is analogous to the potential of a charge density in electrostatics, specifically the charge density 
$\rho(z\,')=z'=\rho(r\,',\theta')=r'\cos\theta'$ for $r'\leq a$ and $\rho(r\,')=0$ for $r'> a$.
To solve the final integral in Eq. \ref{eq:tauz}, we perform a multipole expansion 
of $\psi(r,\theta)$ in terms of Legendre polynomials $P_l(\cos\theta)$,
exploiting the rotational symmetry of $\psi(r,\theta)$ about the $z$ axis \cite{jackson}, leading to
\begin{equation}
\tau_z=\frac{2a^2}{105 D}\label{eq:tauz2}.
\end{equation} 

Now consider the contribution from an additional gradient with zero mean over the volume of the sphere. 
We need to perform the integrals over a non-uniform distribution in Eq. \ref{eq:tauz}. 
Since a gradient along the $z$-axis contributes a factor $\cos\theta$, which leads to the vanishing integral
$\int_0^\pi\sin\theta\cos^3\theta d\theta=0$, we conclude that only the constant background contributes to
the uncertainty in the gradient measurement. Therefore, Eq. \ref{eq:tauz2} for $\tau_z$ remains true even
when an actual gradient is present.
 
The result for $\tau_z$ in Eq. \ref{eq:tauz2} can be used in Eq. \ref{eq:czT3} to obtain the normalized uncertainty
of gradient measurement by the perfectly monitoring sphere
\begin{equation}
\frac{\langle(\delta c_{\vec r})^2\rangle}{(c_0/a)^2}=\frac{3\langle(\delta c_z)^2\rangle}{(c_0/a)^2}=\frac{15}{7\pi Dac_0 T},
\end{equation}
where all three components of the gradient contribute independently. 
Hence, the perfectly monitoring sphere is not only inferior to the perfectly absorbing sphere for concentration
sensing by a factor of $12/5$ in variance ({\it cf.} Eqs. \ref{eq:res_one_c} and \ref{eq:res_two_c}), 
but is also inferior by an even larger factor of $60/7$ for gradient sensing ({\it cf.} Eq. \ref{eq:res_one}).\\


\begin{table*}[t]
\begin{tabular}{c|c|c|c}
               Measurement uncertainty   & Perfect absorber  &  Perfect monitor  & Ratio absorber/monitor \\
\hline

Concentration: $\frac{\langle(\delta c)^2\rangle}{c_0^2}$           & $\frac{1}{4\pi Dac_0T}$ \cite{berg77} & $\frac{3}{5\pi Dac_0T}$  & $\frac{12}{5}${\large$=2.4$}   \\
Gradient: $\frac{\langle(\delta c_{\vec r})^2\rangle}{(c_0/a)^2}$ & $\frac{1}{4\pi Dac_0 T}$ & $\frac{15}{7\pi Dac_0 T}$ & $\frac{60}{7}${\large$\approx 8.6$} \\
\end{tabular}
\caption{\label{tab:formulas} 
Uncertainties in measured concentration and concentration gradient for two idealized cell models: a perfectly absorbing sphere  
(second column) and a perfectly monitoring sphere (third column). Also provided is the ratio of the uncertainties of the
absorber and monitor. 
Parameters: diffusion constant $D$, radius of sphere $a$, averaging time $T$, and average
chemical concentration $c_0$. }
\end{table*}

\section{Comparison with Experiment}

Van Haastert and Postma \cite{haastert07} recently measured
the Chemotactic Index of Dicty cells in a cAMP gradient \cite{haastert07} (Fig. 1, symbols). 
They used a pipette containing different concentrations of cAMP to establish a distance-dependent 
steady-state cAMP gradients. The Chemotactic Index was defined as the distance moved by the cell 
in the direction of the cAMP gradient divided by the total distance moved.
How does the observed chemotactic ability of Dicty compare with the fundamental limits on gradient
sensing set by diffusion? To facilitate comparison to the results of 
van Haastert and Postma \cite{haastert07}, we have calculated the optimal Chemotactic Index
for a cell acting as a perfectly absorbing sphere.

To obtain the optimal Chemotactic Index {\it CI}, we assume that after averaging for a time $T$, 
a cell moves at a constant velocity in the direction of the estimated gradient. If we take
the actual gradient to point in the $z$-direction, then the chemotactic index for
one run $i$ is simply $\cos\theta_i$, where $\theta_i$ is the angle between the true gradient and the
estimated gradient. If the velocity and run time are
the same for each run, leading to a constant run length $l$, then the average Chemotactic Index is given by 
\begin{equation}
CI=\frac{\sum_i^Nl_z}{\sum_i^Nl}=\frac{l\sum_i^N\cos\theta_i}{Nl}=\langle \cos\theta_i\rangle.\label{eq:CI}
\end{equation}

To evaluate $\langle \cos\theta_i\rangle$ for a perfectly absorbing sphere, we use our result (Eq. \ref{eq:res_one})
for the variance of the estimated gradient in each direction, {\it e.g.} 
$\langle(\delta c_{x,y,z})^2\rangle=c_0/(12\pi Da^3T)$. Assuming a 
Gaussian distribution with these variances, as well as an actual gradient with mean value $c_z^0$ 
in the $z$ direction, the 2-dimensional distribution of estimated gradients $c_{\vec r}=(c_x,c_z)$ 
is given by
\begin{equation}
P_{c_z^0}(c_x,c_z)=\frac{1}{2\pi\sigma_x\sigma_z}e^{-c_x^2/(2\sigma_x^2)-(c_z-c_z^0)^2/(2\sigma_z^2)},\label{eq:P}
\end{equation}
where $\sigma_{x,z}=\sqrt{\langle(\delta c_{x,z})^2\rangle}$.
From this distribution, we can obtain the optimal Chemotactic Index
\begin{eqnarray}
CI&=&\langle \cos\theta\rangle=\langle c_z/\sqrt{c_x^2+c_z^2}\rangle\nonumber\\
&=&\sqrt{\frac{\pi y}{2}}e^{-y}[I_0(y)+I_1(y)]\label{eq:CI2},
\end{eqnarray}
where $y=(c_z^0)^2/(4\sigma_z^2)$ and $I_{0(1)}$ is the first (second) order modified Bessel function 
of the first kind. Fig. 1 shows a 
comparison of the optimal {\it CI} (solid curves) with the data of Ref. \cite{haastert07}. Importantly, 
the comparison relies on only a single global fitting parameter representing gradient-sensing ability, namely
the product $Da^3T$ where $D$ is the diffusion constant, $a$ is the cell diameter, and $T$ is the 
averaging time. Based on the estimates $D=300\mu\text{m}^2/$s and $a=5\mu$m \cite{haastert07}, 
the averaging time is predicted to be about $T=3.2$ s. 
(The perfect-monitor model yields an identical curve,
but with a longer inferred averaging time $T=27.5$ s.) The theory for the optimal
Chemotactic Index matches the experiment rather well. 
Eq. \ref{eq:CI2} further predicts that the Chemotactic Index depends on the gradient $c_z^0$ and the
concentration $c_0$ only through the combination $(c_z^0)^2/c_0$ (Fig. 1, inset).
Intuitively, $\sqrt{(c_z^0)^2/c_0}$ measures the signal to noise ratio - the signal
is proportional to the true gradient $|c_z^0|$, while the noise from particle diffusion
scales as $\sqrt{c_0}$. More generally, the optimal Chemotactic Index depends on all 
variables only through the combination $(c_z^0)^2 D a^3 T/c_0$.
Moreover, the theory predicts the full distribution of run angles (see inset of Fig. S1 in supporting information),
which can be obtained by integrating Eq. \ref{eq:P} in the radial direction for each angle $\theta$.


\section{Discussion}

Many types of cells are known to measure spatial chemical gradients directly with high accuracy. 
In particular, {\it Dictyostelium discoideum} (Dicty) is well known to measure extremely 
shallow cAMP gradients important for fruiting body formation \cite{arkowitz99,mato75,haastert07} 
and {\it Saccharomyces cerevisiae} (budding yeast) detects shallow gradients 
of $\alpha$ mating pheromone \cite{segall93}. 
Accurate spatial sensing is also performed by cells of the immune system 
including neutrophils and lymphocytes \cite{zigmond77}.
The question arises what are the fundamental limits 
of gradient sensing set by chemical diffusion? Here we derived such limits using as model cells 
a perfectly absorbing sphere and a perfectly monitoring sphere \cite{berg77}. 
Within the theory, gradients are estimated by comparing the discrete distribution of observed
locations of particles, either absorbed on the surface of the sphere (Fig. 2b) or measured inside 
the sphere (Fig. 2c), with the expected continuous distribution originating from a gradient (Fig. 2a). 
We find that a perfectly absorbing sphere is superior to a perfectly monitoring sphere for concentration 
and gradient sensing by respective factors of 12/5 (= 2.4) and 60/7 ($\approx 8.6$) (Table I), 
since the perfectly absorbing sphere prevents rebinding of already measured particles.
Indeed, the results presented here for the perfectly absorbing sphere represent the true
fundamental limits of both concentration and gradient sensing by cells.
Our theory for the limits of gradient sensing compares favorably with recent measurements by
van Haastert and Postma \cite{haastert07} of Dicty cells migrating to a cAMP-filled pipette (Fig. 1), 
suggesting that Dicty chemotaxis approaches the fundamental limits set by cAMP 
diffusion. 

The marked superiority of the perfect absorber for concentration and gradient sensing leads us to conjecture
that cells may have developed mechanisms to absorb ligands so as to prevent their rebinding. 
Such absorption could be implemented by ligand or ligand-receptor internalization or by degradation of 
bound ligands. Even degradation of ligands without measurement could be advantageous. For example, 
a perfect absorber that measures only a fraction $f$ of incident particles has the same 
uncertainties given in Table I but with an effective measurement time $T_\text{eff}=fT$. Such 
an absorbing cell that measures only 12\% of absorbed particles can still measure gradients 
as accurately as a perfectly monitoring sphere. Similarly, an absorbing sphere of radius $a$ 
with two small measurement patches at its poles of radius $s$ ($s<\!\!<a$), {\it i.e.} with a 
measuring surface-area fraction $s^2/(2a^2)$, effectively reduces the averaging time for pole-to-pole 
gradients to $3s^2T/(2a^2)$ (supporting information). Consequently, a measuring surface fraction 
as small as $4\%$ yields the same uncertainty as the monitoring sphere.  

In fact, there are numerous examples in biology of ligand-receptor internalization \cite{mukherjee97} 
and ligand degradation on cell surfaces, which we speculate, might be related to 
gradient sensing. (1) Although many G-protein-coupled 
receptors are internalized by endocytosis \cite{ferguson01}, the cAMP receptor cAR1 in Dicty is 
not \cite{caterina95}. However, Dicty produces two forms of cyclic nucleotide phosphodiesterase (PDE), 
which degrade external cAMP \cite{malchow72,shapiro83,sucgang97}. One form is membrane-bound (mPDE) and
effectively turns Dicty into an absorber, whereas the other form is soluble (ePDE). 
The membrane-bound form mPDE only accumulates during cell aggregration, 
supporting the idea that degradation of cAMP at the membrane helps 
accurate gradient sensing and navigation. Indeed, cells lacking mPDE display cell-autonomous 
chemotaxis defects even in mixed aggregates with isogenic wild-type cells \cite{sucgang97}.
Interestingly, there is good evidence that Dicty cells do carry out G-protein-coupled
receptor mediated endocytosis of folic acid, another major Dicty chemoattractant \cite{rifkin01}.
(2) In budding yeast, the receptor Ste2 binds $\alpha$-factor pheromone, initiating a mating
response including directed growth (``shmooing") towards a potential mating partner. 
Ligand-bound Ste2 undergoes internalization by endocytosis \cite{schandel94}. 
Furthermore, the protease Bar1 degrades $\alpha$ pheromone externally 
\cite{hicks76,barkai98},
and may be largely membrane associated \cite{ciejek79}.
(3) There are many examples of ligand-receptor internalization in developmental biology. 
For example, primordial germ cells in zebrafish migrate towards the
chemokine SDF-1a that activates the receptor CXCR4b. 
Ligand-induced CXCR4b internalization is required for precise arrival of germ cells at their 
target destination \cite{minina07}. 
These examples suggest a correlation between ligand internalization/degradation
and the accuracy of cell polarization and movement. In presenting these admittedly speculative examples,
our hope is to raise interest, across fields, in how the constraints of
gradient-sensing accuracy may have shaped cellular sensing systems.
 
While the absorption of ligands can improve gradient sensing, there is an inherent problem for 
an absorbing cell to measure a gradient while moving. An absorbing cell moving in a uniform 
concentration creates an apparent gradient due to an 
increased flux of incoming particles at its front and a decreased
flux of particles at its back \cite{berg77}.
Using the model of a spherical cell, the ratio of fluxes between front and back hemispheres is given by
\begin{equation}
R=1+\frac{3av_0}{D}, 
\end{equation}
where $v_0$ is the cell velocity, $a$ is the cell radius, and $D$ is the particle diffusion constant.
On the other hand, the flux ratio of a stationary spherical cell in a gradient $|\vec\nabla c|$ with
uniform background concentration background $c_0$ is given by
\begin{equation}
 R=1+\frac{3a|\vec\nabla c|}{c_0}.
\end{equation}
Hence, a moving cell sees an apparent gradient
\begin{equation}
|\vec\nabla c|=\frac{v_0c_0}{D}.
\end{equation}
As an example, chemotaxis of Dicty to cAMP is observed at a mean concentration of 7nM in a gradient 
of only 0.01 nM/$\mu$m \cite{haastert07}. A Dicty cell moving with a typical speed of 0.2 $\mu$m/s 
at the same mean concentration but without a gradient creates an apparent gradient 
of about half the real gradient. There are several ways out of this dilemma. (1) Cells could separate measurement
from movement at low gradients, {\it e.g.} by stopping, measuring the gradient, and then
moving. Dicty would only need to stop for about $a^2/D\approx 0.1 {\rm s}$ based on 
cell radius of $a=5 \mu {\rm m}$ and diffusion constant $D=300 \mu {\rm m}^2/{\rm s}$. 
(2) Cells could sense gradients transverse to their direction of motion. This
is particularly advantageous for fast moving cells ({\it e.g.} bacteria) for which the apparent gradient
can become more than 100 times steeper than the actual gradient \cite{berg77}. (Indeed, the oxygen-sensing marine 
bacterium {\it Thiovulum majus} directly senses gradients transverse to its direction of motion \cite{thar03}.) 
Interestingly, Dicty cells moving on agar in the absence of a gradient appear to combine these two strategies. 
Qualitatively, the tips of elongated moving cells slow down and flatten, often producing two or more 
distinct pseudopods. Cells then elongate and move ($\sim$ one cell length) in the direction of one of the 
pseudopods before the process is repeated (Liang Li and Ted Cox, personal communication). By this strategy, 
Dicty may avoid locking onto a false, movement-generated gradient. (3) In principle, cells could 
compensate for the apparent motion-generated gradient, either by internal signal processing or by external 
chemical secretion. In fact, Dicty cell do secrete cAMP, primarily from their trailing edge during movement \cite{manahan04}, 
but this cAMP secretion serves dominantly to facilitate cell aggregation, including cells following cells during 
streaming. Given the complex role of cAMP in Dicty aggregation, studies of Dicty chemotaxis using gradients of folate, 
which is absorbed \cite{rifkin01} but apparently not secreted by Dicty, may ultimately prove simpler to interpret.

Our models of the absorbing and the monitoring spheres 
neglect all biochemical reactions, such as particle-receptor binding and downstream signaling,
which could significantly increase measurement uncertainty 
beyond the fundamental limits described here. To study the effects of particle-receptor binding, 
we extended a formalism for the uncertainty of concentration sensing, recently developed by Bialek and Setayeshgar 
\cite{bialek05}, to gradient sensing. We found that the measurement uncertainty
allowing ligand rebinding is larger than the measurement uncertainty without rebinding, confirming
the superiority of the absorber over the monitor (details will be published elsewhere).
A number of mechanistic models for gradient sensing and chemotaxis have addressed the important questions of cell 
polarization, signal amplification, and adaptation \cite{meinhardt99,skupsky05,narang05,levine06,krishnan07,onsum07,otsuji07}, 
cell movement of individual cells \cite{dawes06,dawes07}, cell aggregation with cAMP degradation by PDE \cite{palsson97},
as well as sensing of fluctuating concentrations \cite{bialek05,goodhill99,wylie06}.
Our results on the fundamental limits of gradient sensing complement these models, and
may ultimately help lead to a comprehensive description of eukaryotic chemotaxis \cite{iglesias08}.

Finally, we remark that the experiments by van Haastert and Postma used stationary spatial gradients \cite{haastert07}. 
Cells in such gradients might profit from remembering their direction of motion \cite{andrews07}, and evidence for
such internal memory was recently obtained \cite{li08,samadani06,skupsky07}. It therefore might prove interesting 
to measure the chemotactic index for randomly changing gradients, 
to find out if cells indeed use their memory to improve chemotaxis.

\begin{acknowledgments}
We thank Naama Barkai, John Bonner, Rob Cooper, Ted Cox, Liang Li, Trudi Schupbach, 
Stanislas Shvartsman, and Monika Skoge for helpful suggestions. Both authors acknowledge 
funding from the Human Frontier Science Program (HFSP). RGE acknowledges funding from the
Biotechnology and Biological Sciences Research Council grant
BB/G000131/1 and the Centre for Integrated Systems Biology at Imperial College (CISBIC), NSW
acknowledges funding from National Science Foundation grant PHY-0650617.
\end{acknowledgments}

\end{document}


\title{\Large Accuracy of direct gradient sensing by single cells\\ \ \\
Supplementary information\\ \ \\}

\author{\large Robert G.~Endres$^{1,2}$ and Ned S.~Wingreen$^{1}$\ \\ \ \\}

 \affiliation{
$^{1}$Department of Molecular Biology, Princeton University, Princeton, NJ 08544-1014.\\
$^{2}$Division of Molecular Biosciences, and Centre for Integrated Systems Biology, 
Imperial College, London SW7 2AZ, United Kingdom.\\
}

\maketitle

\section{\label{sec:one}Accuracy of Concentration and Gradient Sensing by a Perfectly Absorbing Sphere}
Consider as a measurement device a sphere of radius $a$ that is a perfect absorber for a certain dissolved
chemical. The device can be used to measure both the local concentration and the local gradient
of the chemical by sensing the chemical current density
$\vec j$ impinging on the surface of the sphere. At steady-state, for the case of a uniform particle
concentration $c_0$ far away from the absorbing sphere, the average current is $J=4\pi aDc_0$, 
where $D$ is the particle diffusion constant \cite{berg77}.
Since the average number of particles $N=4\pi Dac_0T$ absorbed in time $T$ is Poisson distributed, {\it i.e.}
$\langle(\delta N)^2\rangle=\langle N\rangle$,
the perfectly absorbing sphere has a concentration-measurement uncertainty of
\begin{equation}
\frac{\langle(\delta c)^2\rangle}{c^2}=\frac{\langle(\delta N)^2\rangle}{\langle N\rangle^2}
=\frac{1}{4\pi DacT}\label{eq:res_one_c}.
\end{equation}

The perfectly absorbing sphere can also be used to measure the local gradient of the chemical. To calculate the 
current density $\vec{j}$ in this case, we utilize a standard analogy to electrostatics. 
In electrostatics the potential $\phi$ and electric field $\vec E$ in a charge-free environment are 
determined by Laplace's equation, $\nabla^2\phi=0$, and $\vec{E}=-\vec{\nabla}\phi$, respectively. 
As a result, the surface-charge density $\sigma_\text{charge}$ on a conducting sphere (boundary condition 
$\phi=0$ at $r=a$) placed in an electric field of magnitude $E_z$ in the $z$-direction with an additional 
constant potential $\phi$ far away from the sphere is given by the superposition \cite{jackson} 
\begin{equation}
\sigma_\text{charge}=-\left.\frac{1}{4\pi}\frac{\partial \phi}{\partial r}\right|_{r=a}=
-\frac{1}{4\pi}\left(\frac{\phi}{a}-3E_z\cos\theta\right)\label{eq:EM}
\end{equation}
in Gaussian units, where $\theta$ is the polar angle measured with respect to the $z$ axis. 
In the case of a chemical-absorbing sphere, the chemical concentration $c$ 
and the current density $\vec j$ obey equations analogous to the ones governing the potential $\phi$ and
electric field $\vec E$ in electrostatics. Specifically, the spatial dependence of the concentration $c$ follows from the
diffusion equation at steady-state, $\nabla^2c=0$, while the current density is given by $\vec{j}=-D\vec{\nabla}c$. 
Exploiting the result from electrostatics, Eq. \ref{eq:EM},  
the average current density impinging on the perfectly absorbing sphere (boundary condition $c=0$ at $r=a$) in a background 
gradient $c_z=\partial c/\partial z$ in the $z$-direction is given by
\begin{equation}
 j(\theta)=\frac{Dc_0}{a}+3Dc_z\cos\theta\label{eq:Jz}.
\end{equation}  
This expression can be generalized to a gradient $\vec{\nabla} c$ in an arbitrary direction $\vec r$ via
\begin{equation}
 j(\theta,\phi)=\frac{Dc_0}{a}+3D\vec{\nabla} c\cdot \vec e(\theta,\phi)\label{eq:Jr},
\end{equation}  
where $\vec e(\theta,\phi)=(\cos\phi \sin\theta,\sin\phi \sin\theta,\cos\theta)$.

To best estimate the chemical gradient from an observed density of particles absorbed 
at the surface of the sphere during time $T$, we fit the observed density 
$\sigma_T^\text{obs}=\sum_{i=1}^{N}\delta(\vec r-\vec r_i)$, where
$N$ is the total number of absorbed particles, to the expected density $j(\theta,\phi)T$ 
from Eq. \ref{eq:Jr}. The best fit is obtained 
by minimizing the error between the observed density and the expected density
\begin{equation}
 \text{Error}=\int[\sigma_T^\text{obs}-C-\sum_{m=-1,0,1}G_m\,Y_{l=1}^{m}(\theta,\phi)]^2dA,\label{eq:error}
\end{equation}
where $C$ and $G_m$ are the parameters to be determined. Note that in Eq. \ref{eq:error} we write the expected
contribution from the gradient in terms of the spherical harmonics $Y_{l=1}^{m}(\theta,\phi)$: 
$Y_1^{-1}=\sqrt{\frac{3}{8\pi}}\sin\theta e^{-i\phi}$, $Y_1^{0}=\sqrt{\frac{3}{4\pi}}\cos\theta$,
$Y_1^{1}=-\sqrt{\frac{3}{8\pi}}\sin\theta e^{i\phi}$.
Minimizing the error as a function of the parameters $C$ and $G_m$ ($m=-1,0,1$) is achieved by 
setting $\partial \text{Error}/\partial C=0$ and 
$\partial \text{Error}/\partial G_m=0$, which results in the best-fit values
\begin{eqnarray}
 C&=&\frac{\int \sigma_T^\text{obs}\,dA}{\int dA}=\frac{\int  \sigma_T^\text{obs}\,dA}{4\pi a^2}\\
G_{-1}&=&\frac{\int  \sigma_T^\text{obs}Y_{1}^{-1}(\theta,\phi)\,dA}{\int |Y_1^{-1}(\theta,\phi)|^2\,dA}=
\frac{\sqrt{\frac{3}{2\pi}}\int \sigma_T^\text{obs}\sin\theta e^{-i\phi}\,dA}{a^2}\\
G_{0}&=&\frac{\int  \sigma_T^\text{obs}Y_{1}^{0}(\theta,\phi)\,dA}{\int |Y_1^{0}(\theta,\phi)|^2\,dA}=
\frac{\frac{1}{2}\sqrt{\frac{3}{\pi}}\int \sigma_T^\text{obs}\cos\theta\,dA}{a^2}\label{eq:G0}\\
G_{1}&=&\frac{\int  \sigma_T^\text{obs}Y_{1}^{1}(\theta,\phi)\,dA}{\int |Y_1^{1}(\theta,\phi)|^2\,dA}=
\frac{-\sqrt{\frac{3}{2\pi}}\int  \sigma_T^\text{obs}\sin\theta e^{i\phi}\,dA}{a^2}
\end{eqnarray}
and therefore the following best estimates for the background concentration and the individual gradient components:
\begin{eqnarray}
c_0&=&\frac{aC}{DT}\\
c_x&=&\frac{1}{6DT}\sqrt{\frac{3}{2\pi}}(G_{-1}-G_{1})\\
c_y&=&\frac{-i}{6DT}\sqrt{\frac{3}{2\pi}}(G_1+G_{-1})\\
c_z&=&\frac{1}{6DT}\sqrt{\frac{3}{\pi}}G_0\label{eq:cz}.
\end{eqnarray}
Since $c_{x,y,z}$ are estimates of the independent, orthogonal components of the gradient, 
without loss of generality, we consider only the gradient estimate 
in the $z$-direction and generalize to an arbitrary gradient later. From Eq. \ref{eq:cz}, the best
estimate for the gradient in the
$z$-direction after absorption of particles for a time $T$ is given by
\begin{equation}
c_{z}=\frac{\int  \sigma_T^\text{obs}\cos\theta\,dA}{4\pi Da^2T}
=\frac{\sum_{i=1}^{N}\cos\theta_i}{4\pi Da^2T}.\label{eq:gradz}
\end{equation}

We are ultimately interested in the uncertainty (accuracy) of the gradient measurement by the
absorbing sphere, which is given by the variance
\begin{eqnarray}
\langle(\delta c_z)^2\rangle&=&\langle c_z^2\rangle-\langle c_z\rangle^2\\
&=&\frac{\langle\sum_{i=1}^{N}\cos^2\theta_i\rangle+
\langle\sum_{i=1}^{N}\sum_{i\neq j}^{N}\cos\theta_i\cos\theta_j\rangle}{(4\pi Da^2T)^2}-
\frac{\langle\sum_{i=1}^{N}\cos\theta_i\rangle^2}{(4\pi Da^2T)^2}\\
&=&\frac{\langle\sum_{i=1}^{N}\cos^2\theta_i\rangle}{(4\pi Da^2T)^2}\label{eq:dcz_with_1}
=\frac{\langle N\rangle\langle\cos^2\theta\rangle}{(4\pi Da^2T)^2}=\frac{c_0}{12\pi Da^3T}\label{eq:dcz_with_2}.
\end{eqnarray}
Deriving Eq. \ref{eq:dcz_with_2} made use of the independence of the particles to factorize the expectation value
$\langle\sum_{i=1}^{N}\sum_{i\neq j}^{N}\cos\theta_i\cos\theta_j\rangle
=\langle N(N-1)\rangle\langle\cos\theta\rangle^2=\langle N\rangle^2\langle\cos\theta\rangle^2$, 
since $N$ is Poisson distributed. The last step is exact.
We also used $\langle N\rangle=4\pi Dac_0T$, as well as
$\langle\cos^2\theta\rangle=1/3$. The latter relation applied to absorbed particles 
holds even in the presence of a gradient in the $z$ direction 
since the average $\langle\cos^3\theta\rangle$ is zero.

Since the gradient may point in an arbitrary direction, the total uncertainty of the gradient, normalizedby $c_0/a$, 
is given by
\begin{equation}
\frac{\langle(\delta c_{\vec r})^2\rangle}{(c_0/a)^2}=
\frac{3\langle(\delta c_{z})^2\rangle}
{(c_0/a)^2}=\frac{1}{4\pi Dac_0 T}\label{eq:res_one},
\end{equation}
with the factor of 3 arising because each component of the gradient contributes independently to the total uncertainty.
This result for the uncertainty in gradient sensing is independent of the magnitude of the actual gradient present
(including the case when no gradient is present). Interestingly, the result is 
identical to the uncertainty in measureing the concentration by the same perfectly absorbing sphere 
({\it cf.} Eq. \ref{eq:res_one_c}).\\

\noindent{\bf Special case: absorbing sphere with two polar measurement patches}\\ 
We consider a special case in which a whole sphere of radius $a$ is absorbing, but only two small circular patches of radius 
$s\, (<\!\!<a)$ opposite to each other on the sphere measure the number of absorbed particles. 
We start with the analogues of Eqs. \ref{eq:G0} and \ref{eq:gradz} 
for the estimated gradient in the $z$-direction
\begin{eqnarray}
G_0&=&\frac{\int\sigma_T^\text{obs}\cos\theta\,dA}{\sqrt{3\pi}s^2}\\
c_{z}&=&\frac{G_0}{2\sqrt{3\pi}DT}=\frac{\int\sigma_T^\text{obs}\cos\theta\,dA}{6\pi Ds^2T}=\frac{N_1-N_2}{6\pi Ds^2T},
\end{eqnarray}
where we have used $\cos\theta\approx 1$ for patch $1$ and $\cos\theta\approx -1$ for patch $2$.
The variance is given by
\begin{eqnarray}
\langle(\delta c_{z})^2\rangle&=&\langle c_z^2\rangle-\langle c_z\rangle^2\\
&=&\frac{\langle(N_1-N_2)^2\rangle}{(6\pi Ds^2T)^2}-\frac{\langle(N_1-N_2)\rangle^2}{(6\pi Ds^2T)^2}\\
&=&\frac{\langle (\delta N_1)^2\rangle+\langle (\delta N_2)^2\rangle}{(6\pi Ds^2T)^2}\\
&=&\frac{c_0}{18\pi Ds^2aT}
\end{eqnarray}
where we have used $\langle\delta N_i^2\rangle=\langle N_i^2\rangle-\langle N_i\rangle^2=
\langle N_i\rangle$ ($i=1,2$) for a Poisson process, as well as 
$\langle N_1\rangle=\pi s^2DT(c_0+\langle c_z\rangle\,a)/a$ and $\langle N_2\rangle=\pi s^2DT(c_0-\langle c_z\rangle\,a)/a$.
From this we obtain the normalized uncertainty of the gradient in the $z$ direction 
\begin{equation}
\frac{\langle(\delta c_{z})^2\rangle}{(c_0/a)^2}=\frac{1}{12\pi Dac_0T}\cdot\frac{2a^2}{3s^2},
\end{equation}
{\it i.e.} $2a^2/(3s^2)$ times the uncertainty in the gradient using the whole sphere. Hence,
the smaller the patch size $s$, the larger in the uncertainty of the gradient measurement.

\section{Accuracy of Gradient Sensing by a Perfectly Monitoring Sphere}
Berg and Purcell considered a ``perfect instrument'' for measuring the concentration of a certain
dissolved chemical, namely a virtual sphere of radius $a$ that could exactly count the number of diffusing
chemical molecules inside its volume \cite{berg77}. They showed that such an instrument, making 
measurements for a time $T$, could estimate a concentration with an uncertainty 
\begin{equation}
\frac{\langle(\delta c)^2\rangle}{c^2}=\frac{3}{5\pi DacT}\label{eq:res_two_c}.
\end{equation}
We note that the ``perfect instrument'' is actually inferior by a factor 12/5 in variance 
to the perfectly absorbing sphere for concentration measurement (Eq. \ref{eq:res_one_c})
because the absorbing sphere removes particles from the environment, and hence does not measure
the same particle more than once.

We extend Berg and Purcell's analysis of the perfect instrument to include gradient sensing by assuming
that the virtual sphere also measures the positions of all particles in its volume.
As in the previous section, we derive a best estimate for the gradient and the variance of this best
estimate to find the uncertainty of the gradient determination.   
We start by fitting a gradient model $c=c_0+\vec r\cdot\vec c_r$ to the observed time-averaged  
number density $\frac{1}{T}\int dt\rho_\text{obs}(t)=
\frac{1}{T}\int dt\sum_{i=1}^{N}\delta(\vec r-\vec r(t))$, 
obtained by measuring the exact positions of the particles inside the volume of the sphere
for a time $T$. Specifically, we minimize the error
\begin{equation}
\text{Error}=\int\left(\frac{1}{T}\int\rho_\text{obs}(t)dt-c_0-\vec r\cdot\vec c_{r}\right)^2dV.
\end{equation}
As before we focus on one component of the gradient, namely the gradient in the $z$-direction $c_z$. 
By setting $\partial \text{Error}/\partial c_z=0$, we obtain as a best estimate for the $z$ gradient
\begin{equation}
c_{z}=\frac{\frac{1}{T}\int dt\int dV z \rho_\text{obs}(t)}{\int dV z^2}.
\end{equation}
We are interested in the variance of this estimated gradient
\begin{eqnarray}
 \langle(\delta c_z)^2\rangle&=&\langle c_z^2\rangle-\langle c_z\rangle^2\\
 &=&\left(\frac{15}{4\pi a^5}\right)^2\cdot\frac{1}{T^2}\left[\left\langle\left(\int dt\int dV z\rho_\text{obs}(t)\right)^2\right\rangle
 -\left\langle\left(\int dt\int dV z\rho_\text{obs}(t)\right)\right\rangle^2\right]\\
 &=&\left(\frac{15}{4\pi a^5}\right)^2\left[\langle m_{z,T}^2\rangle
 -\langle m_{z,T}\rangle^2\right]\label{eq:czT2},
\end{eqnarray}
where we have used $\int dV z^2=4\pi a^5/15$ and we have defined
\begin{eqnarray}
 \langle m_{z,T}^2\rangle&=&\frac{1}{T^2}\left\langle\left(\int dt\int dV z \rho_\text{obs}(t)\right)^2\right\rangle=
 \frac{1}{T^2}\int_0^T dt\int_0^T dt' \langle m_z(t)m_z(t')\rangle,\label{eq:mzmz}\\
 \langle m_{z,T}\rangle&=&\frac{1}{T}\left\langle\left(\int dt\int dV z \rho_\text{obs}(t)\right)\right\rangle=
 \frac{1}{T}\int_0^T dt\langle m_z(t)\rangle.\label{eq:mzT}
\end{eqnarray}
The quantity $m_z(t)$ is the total of the $z$ coordinates of all the particles inside the sphere at time $t$.
To calculate $m_z(t)$, consider the sphere imbedded inside a much larger volume containing a total of $M$ particles. This results
in $m_z(t)=\sum_{i=1}^{M}z_i(t)$, where $z_i$ is the z-coordinate of particle $i$ if this particle is inside the 
sphere and zero if it is outside. If $N(t)$ is the number of particles inside the sphere at time $t$, then 
on average there will be $\langle N\rangle=\frac{4}{3}\pi a^3 c_0$ particles inside the sphere.
The auto-correlation function $\langle m_z(t)m_z(t')\rangle$ of particles inside the sphere at 
time $t$ and time $t'$ can consequently be calculated
\begin{eqnarray}
 \langle m_z(t)m_z(t')\rangle&=&\langle\sum_{i=1}^M\sum_{j=1}^Mz_i(t)z_j(t')\rangle\\
 &=&\langle\sum_{i=1}^Mz(i)z_i(t')\rangle+\langle\sum_{i=1}^Mz_i(t)\rangle\cdot\langle\sum_{j\neq i}^Mz_j(t')\rangle\\
 &=&\langle N\rangle\,\langle z(t)z(t')\rangle+\langle N\rangle^2\,\langle z(t)\rangle^2\label{eq:mzmz1}\\
 &=&\frac{4}{3}\pi a^3c_0\,u(t-t')+\langle m_z(t)\rangle^2\label{eq:mzmz2},
\end{eqnarray}
where we defined $u(t-t')=\langle z(t)z(t')\rangle$. 
For Eq. \ref{eq:mzmz1} we used that $\langle\sum_{j\neq i}^Mz_j(t')\rangle\approx\langle\sum_{j=1}^Mz_j(t')\rangle$,
which is exact in the thermodynamic limit of the large embedding volume and large $M$. Furthermore, since $N(t)$ and $z(t)$ are
independent, the factorization of the expectation values is exact as well. Plugging Eqs. \ref{eq:mzmz2}
and \ref{eq:mzT} into Eq. \ref{eq:czT2} results in
\begin{equation}
\langle(\delta c_z)^2\rangle=\frac{75c_0}{4\pi a^7T^2}\int_0^T dt\int_0^T dt' u_0(t-t'),\label{eq:dcz2}
\end{equation}
where the second term in Eq. \ref{eq:mzmz2} cancels Eq. \ref{eq:mzT}.

By introducing a correlation time $\tau_z=1/a^2\cdot\int_0^\infty d\tau u(\tau)$ for $z(t)$, the double time integral 
in Eq. \ref{eq:dcz2} can be simplified, provided time $T$ is much larger than $\tau_z$. Using 
time-reversal symmetry $u(\tau)=u(-\tau)$ for equilibrium diffusion (small gradients), the variance transforms into
\begin{equation}
\langle(\delta c_z)^2\rangle=\frac{75 c_0\tau_z}{2 \pi a^5 T}\label{eq:czT3}.
\end{equation}
The remaining task is to calculate $\tau_z$, the probability that a particle with coordinate $z$ inside the sphere at time $t=0$
is still (or again) inside the sphere at a later time $\tau$. 
Since the chemical concentration consists of a constant background concentration plus a gradient, 
we first consider the constant background and later consider the additional gradient.

Based on the solution of the diffusion equation, if a unit amount of chemical is released at 
point $\vec r\,'$, the concentration at point $\vec r$ at a later time $\tau$ is given by
\begin{equation}
f(|\vec r-\vec r\,'|,\tau)=\frac{\exp\left(-\frac{|\vec r-\vec r\,'|^2}{4D\tau}\right)}{(4\pi D\tau)^\frac{3}{2}}.
\end{equation}
Using the result for the time integral in Ref. \cite{berg77}
\begin{equation}
\int_0^\infty d\tau\,f(|\vec r-\vec r\,'|,\tau)=\frac{1}{4\pi D|\vec r-\vec r\,'|},
\end{equation}
the correlation time $\tau_z$ can be expressed as a volume integral over the sphere (the initial coordinate
$\vec r'$ is uniform in the sphere because we assume a uniform background chemical concentration)
\begin{eqnarray}
\tau_z&=&\frac{1}{\frac{4}{3}\pi a^3}\int_0^\infty d\tau\int\,dV\,z\int\,dV'\,z\,'f(|\vec r-\vec r\,'|,\tau)\\
&=&\frac{3}{16\pi^2 Da^3}\int\,dV\,z\int\,dV'\frac{z\,'}{|\vec r-\vec r\,'|}
=\frac{3}{16\pi^2 Da^3}\int\,dV\,z\,\psi(r,\theta)\label{eq:tauz},
\end{eqnarray}
where $r=|\vec r|$. The function
$\psi(r,\theta)$ is analogous to the potential of a charge density that is rotationally symmetric 
around the $z$-axis in electrostatics, {\it i.e.}
$\rho(z\,')=z'=\rho(r\,',\theta')=r'\cos\theta'$ for $r'\leq a$ and $\rho(r\,')=0$ for $r'> a$.
To solve the integral in Eq. \ref{eq:tauz}, we perform a multipole expansion 
of the potential 
\begin{equation}
\psi(r,\theta)=\int dV'\frac{r'\cos\theta'}{|\vec r-\vec r'|}
\end{equation}
in terms of Legendre polynomials $P_l(\cos\theta)$,
exploiting the rotational symmetry about the $z$ axis \cite{jackson}. 
One needs to differentiate two cases:\\

\noindent{\bf Case I ($r'<r\leq a$):}\\
\begin{equation}
\psi_>(r,\theta)=\sum_{l=0}^\infty\left(\frac{Q_l}{r^{l+1}}\right)P_l(\cos\theta),
\end{equation}
where the exterior multipole moments are given by
$Q_l=\int dV'\rho(\vec r\,'){r'}^lP_l(\cos\theta')$ with $\rho(\vec r\,')=r'\cos\theta'$.
Performing the integral, only the dipole moment survives
\begin{equation}
Q_1=\frac{4\pi r^5}{15},
\end{equation}
yielding 
\begin{equation}
\psi_>(r,\theta)=\frac{4\pi r^3}{15}\cos\theta.
\end{equation}
\noindent{\bf Case II ($r<r'\leq a$):}\\
\begin{equation}
\psi_<(r,\theta)=\sum_{l=0}^\infty I_l r^l P_l(\cos\theta),
\end{equation}
where the interior multipole moments are defined as
$I_l=\int dV'\frac{\rho(\vec r')}{{r'}^{l+1}}P_l(\cos\theta')$ and, again, only the dipole moment survives
\begin{equation}
I_1=\frac{2\pi (a^2-r^2)}{3},
\end{equation}
yielding
\begin{equation}
\psi_<(r,\theta)=\frac{2\pi r(a^2-r^2)}{3}\cos\theta.
\end{equation}
Using these expressions for the potential, the remaining integral in Eq. \ref{eq:tauz} can be performed by
summing up the two contributions to the potential $\psi(r,\theta)=\psi_<(r,\theta)+\psi_>(r,\theta)$
\begin{equation}
\tau_z=\frac{2\pi}{\frac{16}{3}\pi^2Da^5}\int_0^a dr\,r^3\int_0^\pi d\theta\, \sin\theta\, \cos\theta 
\ [\psi_<(r,\theta)\ +\ \psi_>(r,\theta)]=\frac{2a^2}{105 D}\label{eq:tauz2}.
\end{equation} 
Now consider the contribution from an additional gradient with zero mean over the volume of the sphere. 
We need to integrate $\int\,d^3\vec r...$ over a non-uniform distribution in Eqs. \ref{eq:tauz} 
and \ref{eq:tauz2}. Since a gradient along the $z$-axis 
contributes a factor $\cos\theta$, leading to the vanishing integral
$\int_0^\pi\sin\theta\cos^3\theta\,d\theta=0$, only the constant background contributes to
the uncertainty in the gradient measurement.
 
The result for $\tau_z$ (Eq. \ref{eq:tauz2}) can be used in Eq. \ref{eq:czT3} to obtain the normalized uncertainty
of the gradient measurement by the perfect instrument
\begin{equation}
\frac{\langle(\delta c_z)^2\rangle}{(c_0/a)^2}=\frac{5}{7\pi Dac_0 T}.
\end{equation}
Since each component of the gradient contributes independently, the total normalized uncertainty is finally
\begin{equation}
\frac{\langle(\delta c_{\vec r})^2\rangle}{(c_0/a)^2}=\frac{15}{7\pi Dac_0 T}\label{eq:final}.
\end{equation}
Hence, the ``perfect instrument'' is not only inferior to the perfectly absorbing sphere for concentration
sensing by a factor of $12/5$ in variance ({\it cf.} Eqs. \ref{eq:res_two_c} and \ref{eq:res_one_c}), 
but is also inferior by an even larger factor of $60/7$ for gradient sensing (Eq. \ref{eq:res_one}).

\newpage

\begin{figure}             
\includegraphics[width=12cm,angle=0]{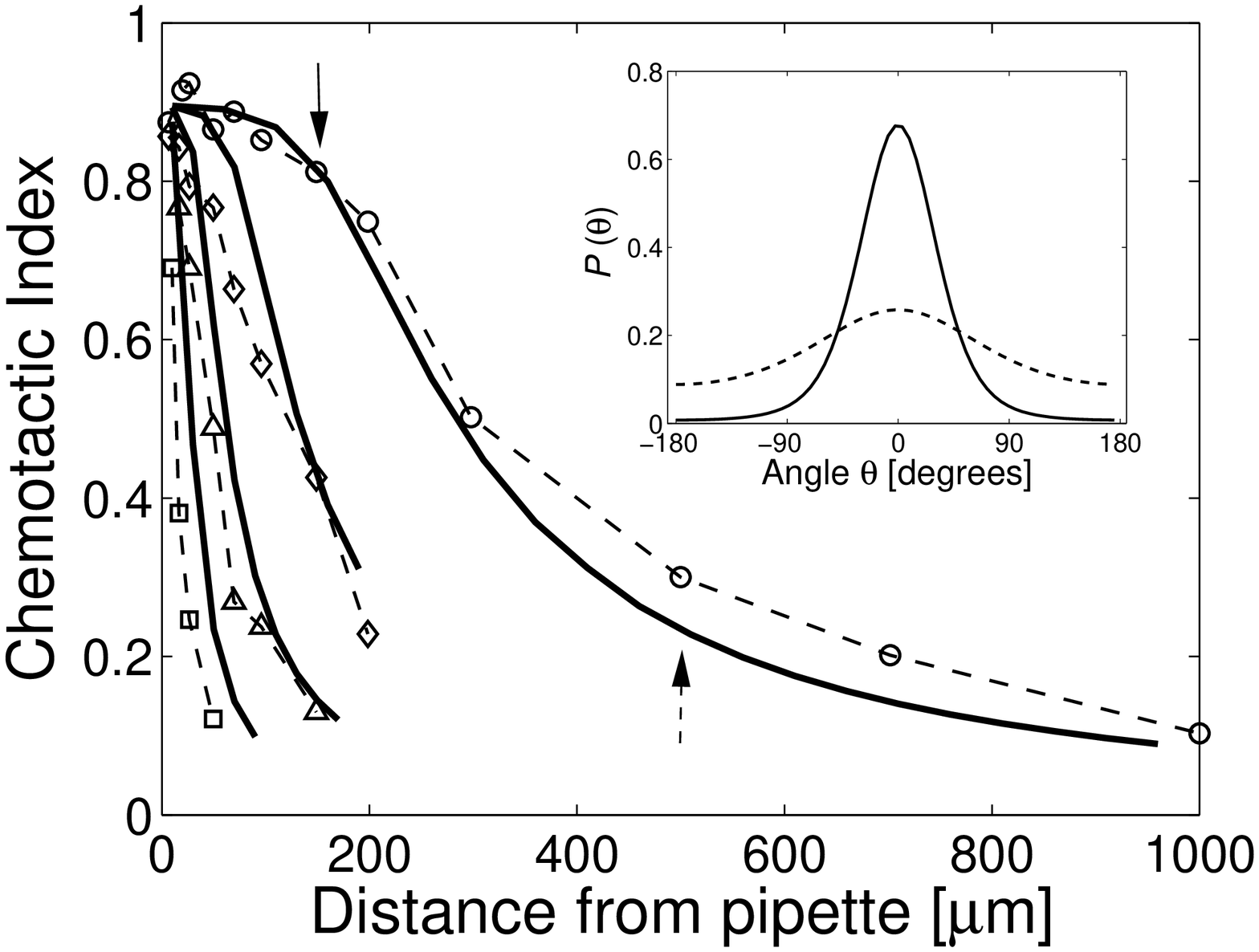}
\caption{Main panel is reproduced from main text Fig. 1, showing the Chemotactic Index of 
{\it Dictyostelium discoideum} cells migrating toward a pipette. Symbols correspond to four different 
cAMP concentrations, 0.1 $\mu$M (squares), 1 $\mu$M (triangles), 10 $\mu$M (diamonds), 100 $\mu$M (circles).  
Inset: predicted distributions of cell-movement directions at the
two points of the 100 $\mu$M cAMP curve indicated by arrows, distances 200 $\mu$m (solid curve and arrow) and 
500 $\mu$m (dashed curve and arrow) from the pipette.}
\label{fig:figS1}
\end{figure}